\documentclass[prl,amssymb,twocolumn,tightenlines,showpacs,superscriptaddress]{revtex4}

\usepackage{graphicx}
\usepackage{amsmath}

\begin{document}

\title{Simple and secure quantum key distribution with biphotons}

\author{I. Bregman}
\affiliation{Dept. of Computer Science, Hebrew University of Jerusalem, Jerusalem 91904, Israel}
\author{D. Aharonov}
\affiliation{Dept. of Computer Science, Hebrew University of Jerusalem, Jerusalem 91904, Israel}
\author{M. Ben-Or}
\affiliation{Dept. of Computer Science, Hebrew University of Jerusalem, Jerusalem 91904, Israel}
\author{H.S. Eisenberg}
\affiliation{Racah Institute of Physics, Hebrew University of Jerusalem, Jerusalem 91904, Israel}

\pacs{03.67.Dd, 42.50.Ex}

\begin{abstract}
The best qubit one-way quantum key distribution (QKD) protocol can
tolerate up to 14.1\% in the error rate. It has been shown how this
rate can be increased by using larger quantum systems. The
polarization state of a biphoton can encode a three level quantum
system - a qutrit. The realization of a QKD system with biphotons
encounters several problems in generating, manipulating and
detecting such photon states. We define those limitations and find
within them a few protocols that perform almost as well as the ideal
qutrit protocol. One advantage is that these protocols can be
implemented with minor modifications into existing single photon
systems. The security of one protocol is proved for the most general
coherent attacks and the largest acceptable error rate for this
protocol is found to be around 17.7\%. This is the first time, to
the best of our knowledge, that the security of qutrit QKD protocols
is rigorously analyzed against general attacks.
\end{abstract}

\maketitle

In order to establish an unconditionally secure communication
channel between two parties (traditionally called Alice and Bob),
they have to share a random sequence of bits known only to them - a
one time pad. Quantum key distribution (QKD) is a scheme that
exploits the details of quantum measurements for generating such a
key. In the most basic protocol (referred to as BB84 after its
inventors Bennett and Brassard\cite{BB84}), Alice sends Bob a series
of two level quantum systems, referred to as qubits. The states are
randomly chosen from two sets, each of them contains two orthogonal
states that represent the logical zero and one. The two sets relate
to each other in a mutually-unbiased way, i.e. the probability of
measuring any particular state when given a state from a different
set is 1/2. Thus, if Alice chooses to send a state from the first
set, a measurement in the basis of the other, either by Bob or by an
eavesdropper (called Eve), will give no information about Alice's
choice.

To create the required secret key, a few more steps should be
carried out through a classical channel, not necessarily secured.
First, Alice and Bob compare their measurement bases and sift only
those bits which were measured in identical bases. From the
remaining key, they reveal a portion and compare the results in
order to estimate the noise parameter. This noise can result either
from a real physical noise in the channel as well as from Eve's
measurements. Next, they perform two transformations on the key, one
to correct for errors and the second, called privacy amplification,
to reduce the amount of mutual information between them and Eve.

A QKD protocol is characterized by a few parameters. The ratio
between the number of remaining bits after completing this procedure
to the number of bits before it, is called the rate of the protocol.
The merit function that characterizes a specific protocol shows its
rate as a function of the disturbance, the error probability that
the channel and Eve have created. The higher the critical
disturbance, where the rate approaches zero, the more useful is the
protocol. In recent years, the lower bound on the critical error
rate of BB84 was improved several times and the best known result is
about $12.4\%$\cite{Renner}.

The BB84 protocol can be extended in many ways. The first simple way
is to add an extra base that is mutually-unbiased with both
others\cite{Bruss}. It can be easily shown that only one mutually
unbiased base (MUB) can be added to the BB84 protocol. According to
Ref. \cite{Renner}, the current critical error rate of this three
MUB protocol is around $14.1\%$. Another approach is to use quantum
systems of higher dimensionality\cite{Peres}. A three level system
can represent a quantum trit (qutrit) and a general $d$-level system
can represent a qudit. A possible advantage is the larger number of
MUB, up to $d+1$ for a $d$-dimensional protocol\cite{Wootters}.
Previously, generalized protocols that use qutrits and higher
dimensional quantum systems have been suggested and their security
was studied\cite{Peres,Karimipour}. A potential advantage of
improved rates for such protocols was shown by examining various
attack
schemes\cite{Tittel,Bourennane,Macchiavello,Bourennaneet,Cerf,Caruso}.
However, to the best of our knowledge, the exact values of the
critical error rates for qutrit protocols subject to general attacks
have never been calculated prior to this work.

There are a few approaches for the realization of more than two
level quantum systems with light. One is by discriminating between
modes of different orbital angular momentum\cite{Mair,Groblacher}.
Another approach is to use several spatial modes\cite{Bartuskova} or
time bins\cite{Thew}. A different approach that has been studied
extensively in recent years is the qutrit representation of the
polarization state of two indistinguishable photons - a
biphoton\cite{Burlakov,Chekhova}. A general biphoton state can be
written as
\begin{equation}\label{GeneralPsi}
|\psi_2\rangle=\alpha_0|2,0\rangle+\alpha_1|1,1\rangle+\alpha_2|0,2\rangle\,,
\end{equation}
where $|n_h,n_v\rangle$ is a Fock representation of $n_h$ ($n_v$)
horizontally (vertically) polarized photons. The general state
$\psi_2$ is represented by a complex vector
$\bar{\alpha}=(\alpha_0,\alpha_1,\alpha_2)$ with four degrees of
freedom (3 complex numbers less a general phase and normalization).
This scheme can be also extended to higher dimensions by adding more
photons, but in this Letter we focus on qutrits.

There are a few difficulties with generating arbitrary biphoton
states\cite{Bogdanov} as well as when trying to
manipulate\cite{Lanyon} and detect them. Optical parametric
down-conversion (PDC) is the obvious choice as a generating scheme,
but there are only two types of processes available; type I that
creates states of the $|2,0\rangle$ and $|0,2\rangle$ forms, and
type II that creates $|1,1\rangle$ (both in a collinear scheme). If
a non-collinear scheme is used, type I can create the $|1,1\rangle$
type as well. In order to create a general state as in Eq.
\ref{GeneralPsi}, a sensitive interferometer that includes both type
I and II crystals is required. Moreover, as we will show later, it
is impossible to transform efficiently by means of linear optics a
state created with one PDC process into an arbitrary biphoton state.
Finally, as transformations are limited, efficient detection is only
possible for the three basic vectors (the components of Eq.
\ref{GeneralPsi}, defined as the 'measurement basis') and their
available transformations within those limits. Detection of a
general state is only possible with a beam-splitter setup that
detects a general state only 1 out of 4 tries.

In this Letter, we define the subset of biphoton states which is
easily generated, manipulated and detected. We find biphoton QKD
protocols within these limits that are more secure and efficient
than the best single photon protocol. Critical error rates are
derived for various qutrit protocols subject to general attacks, and
compared to the best possible qutrit protocol of 4 MUB\cite{Peres}.

First, we shall define the set of allowed transformations. When a
biphoton is transmitted through a linear optics setup, the
polarization state of both of its photons experience the same single
photon unitary transformation. The set of all possible unitary
transformations on a single photon polarization can be mapped onto
the Poincar\'{e} sphere (or the Bloch sphere for a general qubit
realization), such that $\hat{U}(\theta,\varphi)$ describes the
operation that transforms the state at the north pole to the
coordinates $(\theta,\varphi)$. Thus, if we position the state
$|1,0\rangle$ at the north pole, a general operation will transform
it into
\begin{equation}\label{GeneralU}
\hat{U}(\theta,\varphi)|1,0\rangle=\cos(\theta/2)|1,0\rangle+\sin(\theta/2)e^{i\varphi}|0,1\rangle\,.
\end{equation}
We name the north pole state as the 'anchor' state and the set of
$\hat{U}$ operators as the single photon operations. In this single
photon case it is trivial to show that whatever anchor state is
chosen, the $\hat{U}$ operators will always cover the whole single
photon polarization space. There are a few simple rules to note. A
trivial $2\pi$ rotation along any great circle will bring any state
to itself, while a $\pi$ rotation will transform any state to its
orthogonal state. Moreover, a rotation by only $\pi/2$ along such a
path always transforms between two mutually unbiased states.

We shall now find the set of states that can be reached by applying
$\hat{U}$ to the biphoton measurement basis. Choosing the states
$|2,0\rangle$ and $|1,1\rangle$ as the anchors and applying the
single photon operations, we get
\begin{eqnarray}
\label{Sphere}
\hat{U}(\theta,\varphi)(1,0,0)=(\cos^2(\frac{\theta}{2}),\frac{1}{\sqrt{2}}\sin(\theta)e^{i\varphi},\sin^2(\frac{\theta}{2})e^{2i\varphi})\,,\\
\label{Dome}
\hat{U}(\theta,\varphi)(0,1,0)=(\frac{1}{\sqrt{2}}\sin(\theta),\cos(\theta)e^{i\varphi},\frac{-1}{\sqrt{2}}\sin(\theta)e^{2i\varphi})\,,
\end{eqnarray}
as the new $\bar{\alpha}$ vectors, respectively. By allowing only
two parameter operations from a finite set of anchor states, we
defined a two dimensional subset of the general biphoton four
dimensional space. As the $|0,2\rangle$ state appears at the south
pole of the sphere defined by the $|2,0\rangle$ anchor, there is a
full overlap between spheres defined by these two states. On the
other hand, the $|1,1\rangle$ state does not appear on the
$|2,0\rangle$ sphere, thus defining a non-overlapping sphere that
can not be reached from the $|2,0\rangle$ sphere by single photon
operations.

\begin{figure}[tbp]
\includegraphics[angle=0,width=70mm]{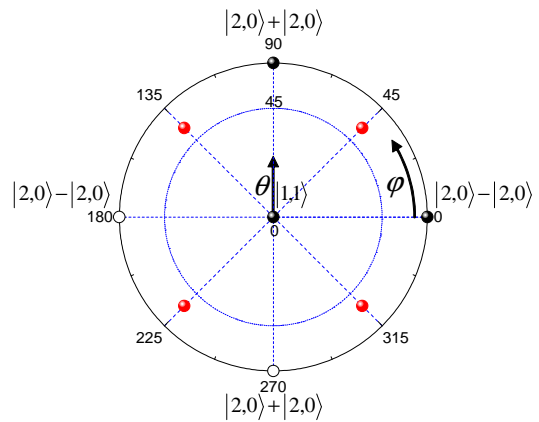}
\caption{\label{fig1}A top projection from the north pole of an
orthogonal triplet basis on the $|1,1\rangle$ dome. The empty
circles are states identical to their $180^\circ$ opposites. Four
states on this manifold are mutually-unbiased to the triplet. These
four states are not orthogonal between themselves.}
\end{figure}
It is possible to identify a few great circle rotation rules for the
two new spheres, that would be useful later when we will look for
possible QKD protocols in this subset. First, the $|2,0\rangle$
sphere inherited all of the properties from the $|1,0\rangle$ single
photon sphere. A $2\pi$ rotation returns to the original state, a
$\pi$ rotation transforms between two orthogonal states and $\pi/2$
between two states whose projection on each other is 1/2. In the
case of the $|1,1\rangle$ sphere, there is an interesting
difference. The same rules still apply, but for half the angles. A
$\pi$ rotation returns to the original state, a $\pi/2$ rotation
transforms between two orthogonal states and $\pi/4$ between two
states whose projection on each other is $1/\sqrt{2}$. We can
identify an orthogonal vector triplet on the $|1,1\rangle$ sphere as
three states with $\pi/2$ in between them. A simple example for such
a triplet is the $|1,1\rangle$ state and two 'bunched' states such
as $|2,0\rangle+|2,0\rangle$ and
$|2,0\rangle-|2,0\rangle$\cite{Hong}. Curiously, the state which is
exactly in the middle of any such triplet (at the tetrahedral point,
about $54.7^\circ$ from all the triplet states) is mutually unbiased
with them (see Fig. \ref{fig1}). As opposite points on the
$|1,1\rangle$ sphere are identical, it is enough to consider only
the upper half of the sphere (a dome).

The rotation rules convert the task of finding MUB to a packaging
problem. Is it possible to pack two triplet bases together on the
$|1,1\rangle$ dome and preserve mutually-unbiased relationship? By
examining Fig. \ref{fig1} it is easy to see that it is impossible.
On the other hand, we recognize a second general type of an
orthogonal basis. If we position the dome concentrically inside the
$|2,0\rangle$ sphere such that both of their north poles point in
the same direction, every ray that passes through the sphere center
will cut the subspace at three points, two on the sphere and one on
the dome. This ray triplet contains three orthogonal states. The
simplest example is the vertical ray that defines the measurement
basis.

\begin{figure}[tbp]
$\begin{array}{c@{\hspace{1in}}c}
\includegraphics[angle=0,width=43mm]{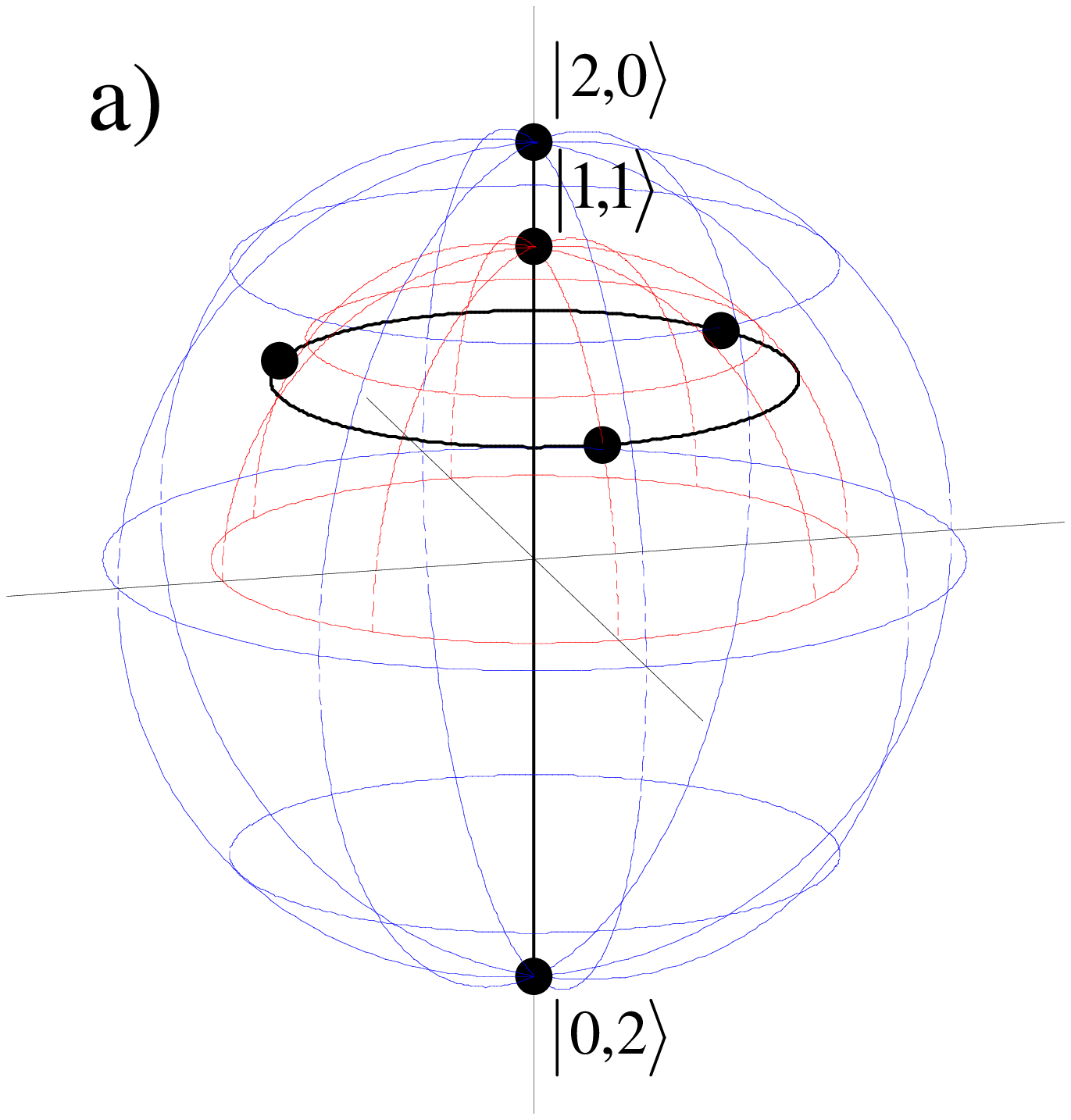}
\includegraphics[angle=0,width=43mm]{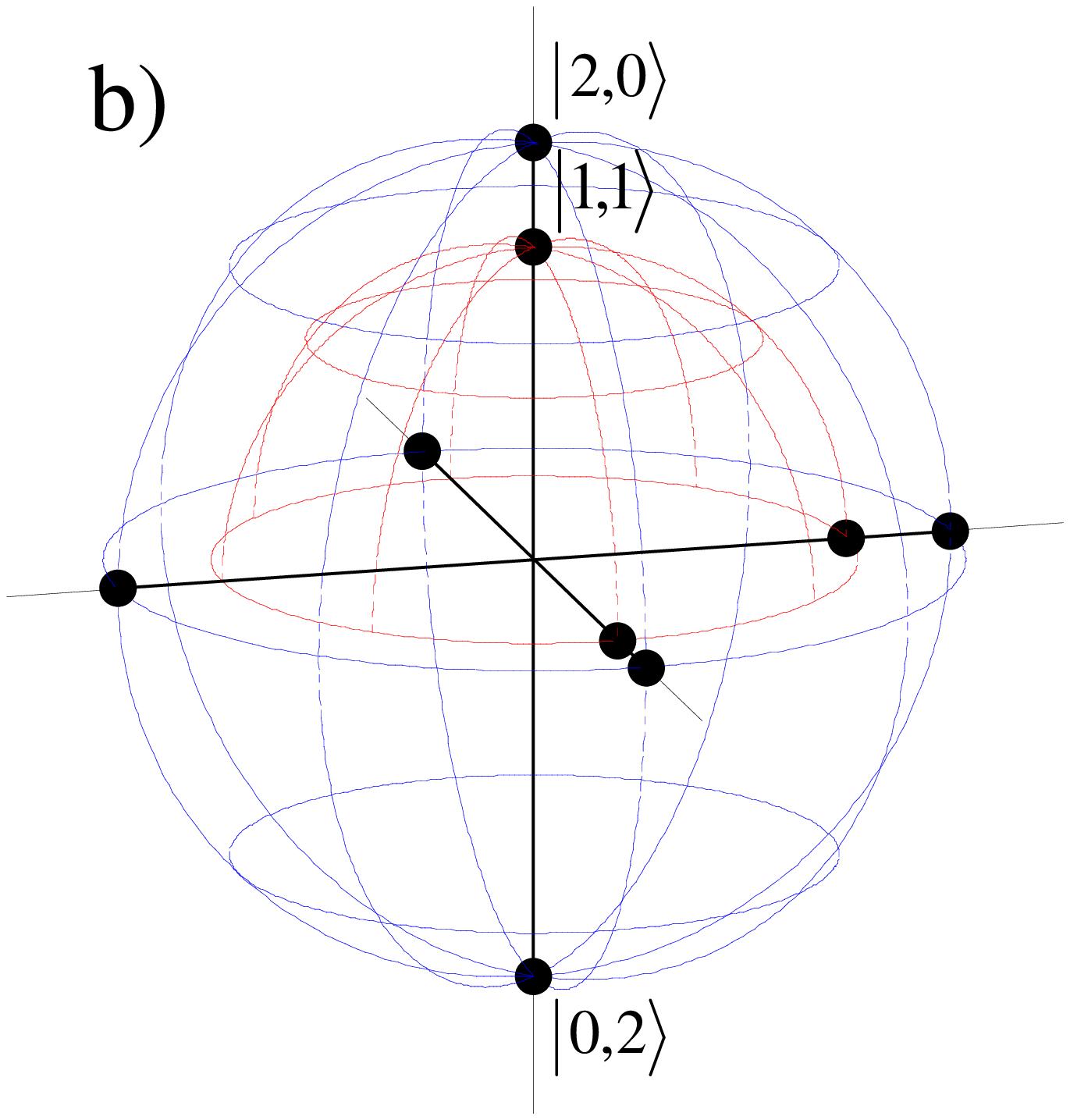}
\end{array}$
\caption{\label{fig2}a) The umbrella protocol: two
mutually-unbiased bases within the single-photon operation
subspace. b) The three rays protocol: three ray-type bases that
are not perfectly mutually-unbiased.}
\end{figure}
We define two possible protocols whose security will be checked
here. The first protocol has two perfectly MUB and the second has
three (or more) which are not. Apart from the measurement basis
which is included in both protocols, the additional basis of the
first protocol is
\begin{equation}\label{Umbrella}
\left\{\frac{1}{\sqrt{3}}(1,1,-1),\frac{1}{\sqrt{3}}(1,\tau,-\tau^2),\frac{1}{\sqrt{3}}(1,\tau^2,-\tau)\right\}\,,
\end{equation}
where $\tau=e^{i\frac{2}{3}\pi}$. Notice how this basis is
equivalent to the second Fourier basis in Ref. \cite{Peres}, up to a
minus sign at the last position. We name this protocol after its
umbrella shape (see Fig. \ref{fig2}a). The two additional bases of
the second protocol are of the ray type:
\begin{eqnarray}\label{ThreeRays}
\left\{\frac{1}{2}(1,\sqrt{2},1),\frac{1}{\sqrt{2}}(1,0,-1),\frac{1}{2}(1,-\sqrt{2},1)\right\}\,,\\
\nonumber
\left\{\frac{1}{2}(1,\sqrt{2}i,-1),\frac{1}{\sqrt{2}}(1,0,1),\frac{1}{2}(-1,\sqrt{2}i,1)\right\}\,.
\end{eqnarray}
Although the three bases are only close to mutually-unbiased, this
protocol is appealing because of its symmetry and similarity to the
three bases protocol for qubits\cite{Bruss} (see Fig. \ref{fig2}b).
Additionally, we considered a protocol with 7 ray bases in order to
check whether adding rays improves the protocol. The 7 rays are the
three of the previous protocol, plus the four rays through the
tetrahedral points (the directions are identical to the 7 solid
circles in Fig. \ref{fig1})

\begin{figure}[tbp]
\includegraphics[angle=0,width=86mm]{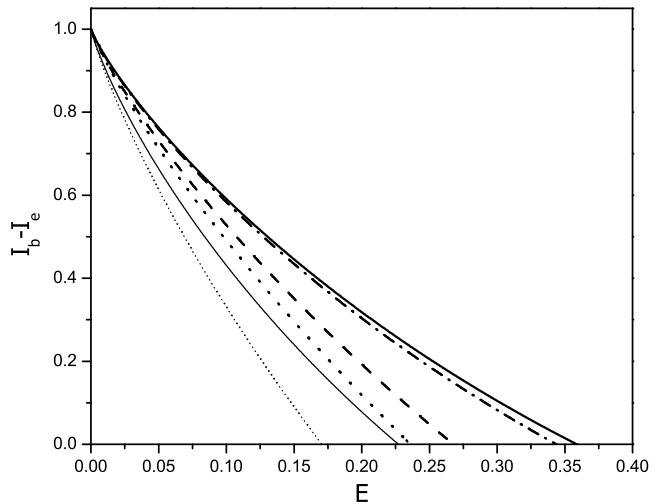}
\caption{\label{fig3}Intercept and resend analysis of a few QKD
protocols. Qubit protocols are in thin lines - dotted for BB84 and
solid for the 3 MUB extension. Thick lines are for qutrit protocols
- dotted, dashed, dot-dashed and solid for the umbrella, 3 rays, 7
rays and 4 MUB, respectively.}
\end{figure}
The simplest eavesdropping attack scheme is the 'intercept and
resend' approach. Showing that a protocol is secure against this
limited attack does not imply security against general attacks. It
is only used here for comparison between different protocols. We
calculated Bob and Eve mutual information with Alice with the method
of Ref. \cite{Peres} and plotted the difference between them as a
function of the error rate for a few protocols (see Fig.
\ref{fig3}). There is a clear hierarchy between qubit and qutrit
protocols. The umbrella protocol is better than any other using
qubits, while the 3 rays protocol is even better than the umbrella
protocol even though it does not include perfectly MUB. As expected,
the best performance belongs to the 4 MUB protocol, that marks the
upper limit for any qutrit protocol. Surprisingly, the 7 rays
protocol performs very close to ideal. All critical values are much
higher than the real limits as this attack scheme is far from ideal.

In order to prove the ultimate security that is required from a QKD
protocol, we use the general method introduced in Ref.
\cite{Renner,Renner2}. The advantage of this method over other
previous security proofs is its easy extendability to higher
dimension while proving security against the most general coherent
attacks. The scheme results in a convex nonlinear optimization
problem for every error rate value. Here, we only prove
unconditional security for the 4 MUB and the umbrella protocols as
they correspond to simpler sets of constraints than the ray
protocols. Nevertheless, it is only reasonable to assume that as the
performance of the ray type protocols according to intercept and
resend analysis is between the 4 MUB and umbrella protocols, their
performance against general coherent attacks will be in this range
as well.

\begin{figure}[tbp]
\includegraphics[angle=0,width=86mm]{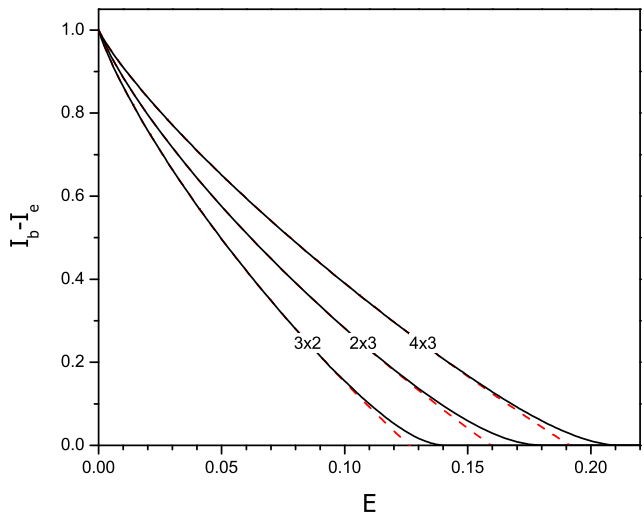}
\caption{\label{fig4}Key generation rate analysis for a general
coherent attack. Solid (dashed) lines are for calculation with
(without) noise optimization. Presented results are for 3 qubit and
2 and 4 qutrit MUB (3x2, 2x3 and 4x3 respectively).}
\end{figure}
After deriving the constraints, we optimized numerically the target
function (the protocol rate) with the CVXOPT convex optimization
package\cite{cvxopt}. Both procedures, with and without noise
addition, were calculated. The results are plotted in figure
\ref{fig4}. For the 3 MUB qubit protocol we reproduced the known
critical values of 14.1\% (12.7\%) (with and without noise
optimization). We find proven lower bounds of 17.7\% (16.0\%),
20.3\% (18.25\%) and 21.1\% (19.1\%) for the critical error rates of
the 2 (the umbrella), 3 and 4 MUB qutrit protocols, respectively.
The 3 qutrit MUB graph is not presented here as it can't be realized
within the borders of our subset. These values are considerably
higher than the best value for qubit protocols to date. Just two MUB
gives most of the gain between qubits and qutrits, as previously
suggested by weaker security analysis\cite{Cerf,Caruso}.

We have left open an important issue regarding any qutrit protocol
with biphotons. Namely, what is the relation between the single
photon (qubit) error rate of a certain channel and the qutrit error
probability when transmitting biphotons? This issue will be
addressed in a later work.

In conclusion, we defined the single photon operation subspace of
the polarized biphoton representation of qutrits. This subspace
includes states which are easy to generate and detect, and thus are
easy to implement in a QKD protocol. We suggested a few possible
protocols within this subspace. The security of these protocols was
analyzed and compared to standard qubit protocols. A large
improvement was shown compared to qubits, even for the umbrella
protocol, which has only two MUB. The unconditional security of the
umbrella and the 4 MUB protocols was proved by extending a previous
proof for qubits. This is the first time, to the best of our
knowledge, that qutrit protocols have received such a rigorous
treatment.

H.S.E. and I.B. would like to thank the Israeli Science Foundation
for supporting this work under grants 366/06 and 1057/05,
respectively. The authors wish to thank B. Kol and N. Gisin for
helpful discussions.

\end{document}